\documentclass[prl,aps,twocolumn,showpacs,amsmath,amssymb]{revtex4}
\usepackage{graphicx}
\usepackage{dcolumn}

\begin{document}

\title{
Pb-graphene-Pb Josephson junctions: characterization in B field.
}

\author{I. V. Borzenets, U.C. Coskun, H. Mebrahtu, and G. Finkelstein}

\affiliation{Department of Physics, Duke University, Durham, NC 27708}

\begin{abstract}
We fabricate superconductor-graphene-superconductor Josephson junctions with superconducting regions made of lead (Pb). The critical current through grapehene may be modulated by external magnetic field; the resulting Fraunhofer interference pattern shows several periods of oscillations, indicating that the junction is uniform. Deviations from the perfect Fraunhofer pattern are observed, and their cause is explained by a simulation that takes into account the sample design. 
\end{abstract}

\maketitle

The properties of superconductor-graphene-superconductor (SGS) junctions have attracted significant attention \cite{heersche_2007,miao_2007,du_2008,gueron_2009, ivan}. Unlike the conventional superconductor-normal\ metal-superconductor (SNS) junctions, devices made with graphene allow for high tunability with the gate voltage. We have recently reported on SGS junctions which use lead (Pb) as the superconducting material \cite{ivan}. Lead has a relatively high critical temperature of 7.2 K, compared to 1.2 K in the case of the commonly used aluminum. Indeed, we have observed supercurrent through graphene at temperatures as high as 2 K. In this paper, we characterize the properties of these junctions by applying magnetic field. 

We fabricate the superconducting contacts to graphene from a Palladium/Lead (Pd/Pb) bilayer. First, we deposit a 2 nm layer of palladium, which creates transparent contacts to graphene \cite{huard_2008,avouris};  100nm layer of lead is deposited in situ on top. The lateral width of the contacts is typically 500 nm. In this work we present the results measured on a junction about $20 \mu$m wide and 400 nm long. (Commonly, length of the junction is defined as the distance between the superconducting contacts, and width as the distance of the normal metal along the superconducting contacts.) In order to create such a wide junction, the leads are bent in two places to fit on a moderately-sized graphene flake (Fig. 1a). We show that this particular sample design has certain nontrivial consequences. 


\begin{figure}[htp]
\includegraphics[width=0.6 \columnwidth]{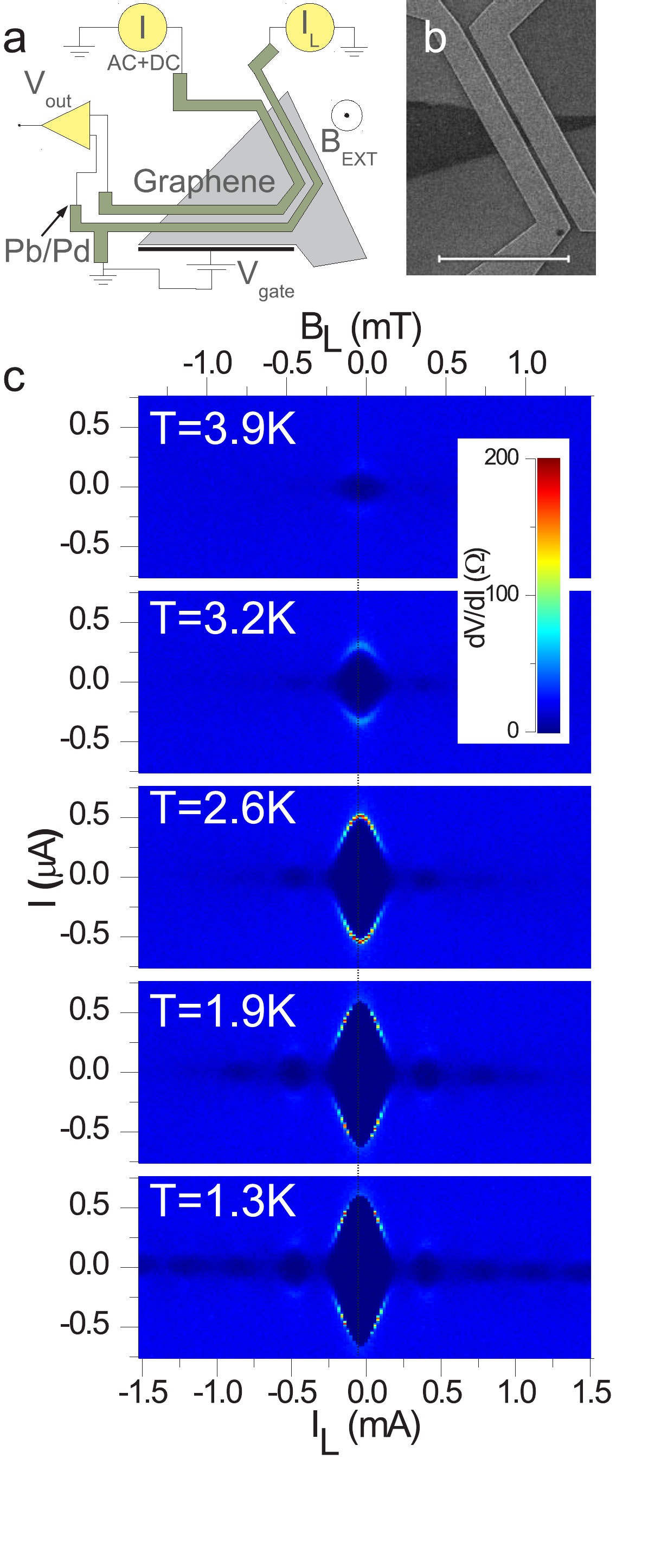}
\caption{\label{fig:overview}
b)  Scanning electron micrograph of a graphene-based SNS Josephson Junction similar but smaller than the one presented in this paper. (Junctions used for measurement were not imaged in order to preserve the quality of graphene.) The dark triangular area is a single layer graphene flake, and the metal contacts are made from lead (Pd) with a thin contact layer of palladium (Pd).  a) Schematic of the measurement setup. The metal leads form a "$\bigsqcup$" shape in order to  increase their length. Bias current $I$ is sent through the junction with a small AC  component. The resulting AC component of the voltage across the junction is measured using a lock-in amplifier allowing one to record the differential resistance $R\equiv dV/dI$. An external magnetic field $B_{ext}$ is applied by a superconducting solenoid. In addition, a magnetic field $B_L$ is created by sending a current $I_L$ along one of the Pb leads of the junction. Sweeping the current $I_L$ allows to apply a very small magnetic field $B_L$. c) Differential resistance $dV/d$I maps measured {\emph vs.} bias current $I$ and magnetic field-inducing current $I_L$. Regions of vanishing $R$ appear dark. Each panel corresponds to measurement at a different temperature. Enhanced zero-bias conductance develops around $\approx 4K$ for small fields $B_L$. With lower temperatures more and more critical current modulations appear and the Fraunhofer interference pattern is observed. At the base temperature of $1.3K$, critical current is seen at fields beyond $5mT$ (only fields up to $1.4mT$ are shown). Observing many oscillations suggest that the junction is highly uniform.   
}
\end{figure}

We measured the sample electronic properties using a pseudo 4-probe setup (Fig. 1a). The junction is biased by current $I$ which contains a small AC component, and the AC voltage across the junction is measured using a  lock-in amplifier. The carrier density in graphene can be tuned by the back-gate voltage $V_{gate}$ but for the results presented the gate voltage is set at zero. Finally, a perpendicular magnetic field can be applied using two methods. Conventionally, a field $B_{ext}$ can be created by an external solenoid magnet. Alternatively, we send a current $I_{L}$ along one of the superconducting leads (Fig. 1a), inducing a field which we label as $B_L$. The advantage of the second method is that the required small fields can be easily obtained and rapidly changed. In this sample we have calibrated $B_{L}$ to be equal to $0.95 \mathrm{\frac {T}{A}} I_{L}$ (see details below).

The Pd/Pb electrodes become superconducting at a temperature of $\approx 7$ K, and the SGS junctions begin to exhibit enhanced zero-bias conductance at temperatures of $\approx 5$ K. Below $\approx 2$ K, a fully formed supercurrent branch is clearly observed \cite{ivan}. Figure 1c demonstrates the differential conductance $R\equiv dV/dI$ versus bias current $I$ (vertical axis) and magnetic field $B_L$ (horizontal axis) measured at several temperatures. The dark areas of the maps in Figure 1c correspond to the regions of suppressed resistance. The regions are bound by a critical current $I=I_C$, above which the junction becomes normal. The value of $I_C$ increases as temperature is lowered and saturates around $I_C\approx 0.5\mu$A at zero magnetic field (see the lowest map in Fig. 1c). When $B_L$ is applied, $I_C$ oscillates in a way closely resembling the Fraunhoffer diffraction pattern \cite{tinkham_1996}. Several oscillations of $I_C$ can be observed at the lowest temperature; this indicates that the junction is uniform. 

We next apply an external magnetic field $B_{ext}$, which is found to shift the modulation pattern of Figure 1c in the horizontal direction (Fig. 2). The shift is linear in $B_{ext}$: indeed, at the center of the pattern the external field and the one induced by $I_{L}$ cancel each other. The observed rate of shift allows us to fix the conversion $B_{L} = 0.95 \mathrm{\frac {T}{A}} I_{L}$ mentioned earlier. This factor is also consistent with our order of magniture estimates. Furthermore, the shift of the pattern by $\Phi_0$ in an external magnetic field of $0.36$ mT allows us to extract the effective area of $5.6 \mathrm{\mu m}^2$. While this area is smaller than the $8 \mu m^2$ expected from the designed sample dimensions of $W=20\mu m$ by $L=0.4\mu m$, it is quite likely that the length $L$ between the leads is reduced in the process of lithography, or that the magnetic field is modified due to the presence of the superconducting leads.

\begin{figure}[t]
\includegraphics[width=0.6 \columnwidth]{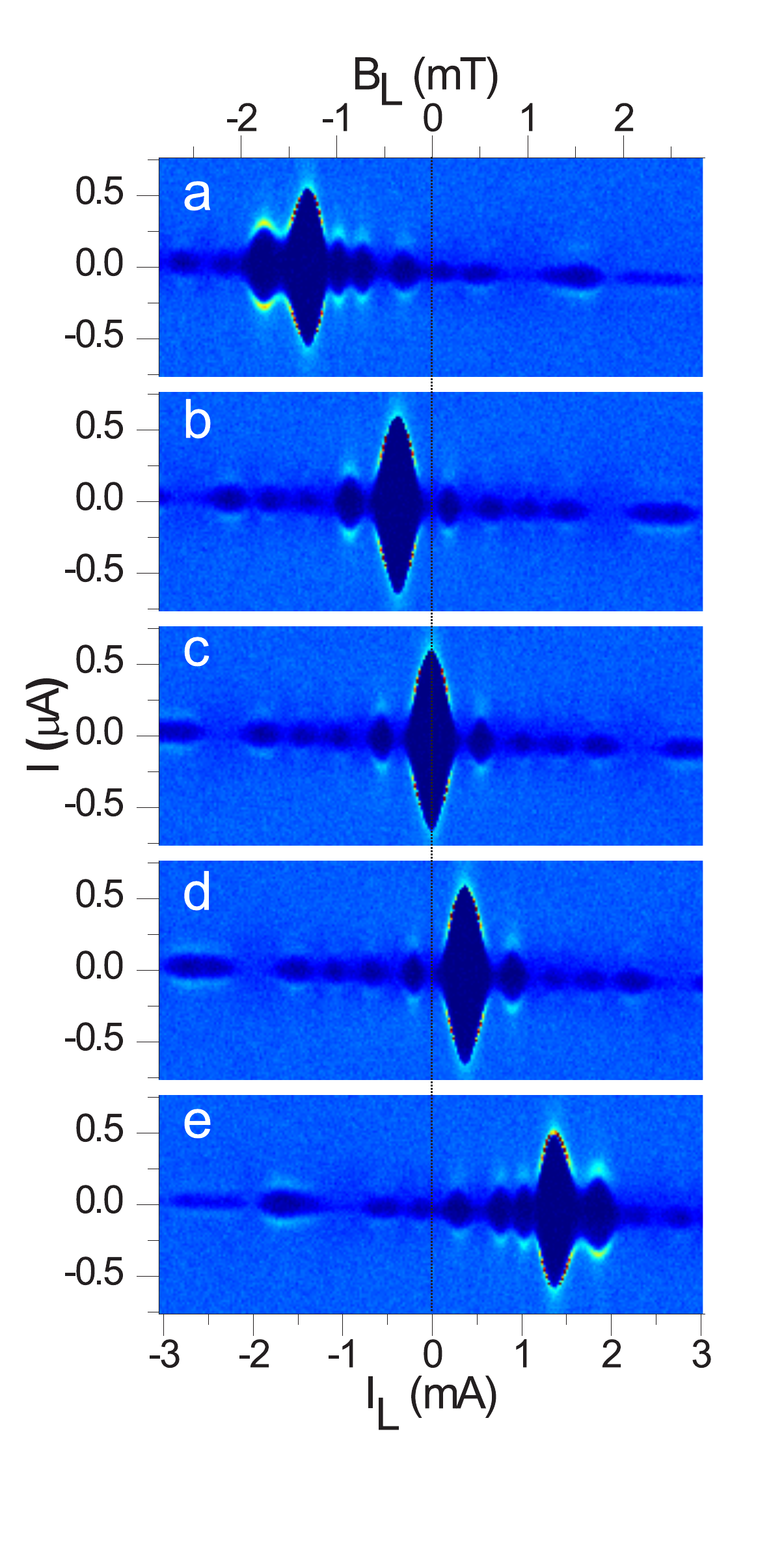}
\caption{\label{fig:FRAUNHOFER2}
$R(I,I_L)$ maps (like those in Fig. 1c) measured at different values of the external magnetic field (a): $B_{ext}= - 1.3$ mT, (b): $B_{ext}= - 0.36$ mT, (c): $B_{ext}=0$, (d): $B_{ext}=0.36$ mT, (e): $B_{ext}= 1.3$ mT. Application of $B_{ext}$ shifts the modulation pattern, so that at its center $B_{ext}$ and $B_L$ cancel each other (b-d). Since the cancellation is not perfect, the pattern gets distorted (a), compared to the pattern at zero external field (c). An opposite orientation of the external magnetic field (e) results in mirror reversal of the distortions. $T = 1.3$ K.
}
\end{figure}

When magnetic field $B_{ext}$ of the order of tens of mT is applied to the sample, the observed pattern becomes distorted even after the field is returned back to zero (Fig. 3). It is clear that the resulting pattern at $B_{ext}=0$ (Fig. 3a) is very different from the original one (Fig. 2c). We can partially recover the original pattern by setting $B_{ext}\approx 3.4$mT (Fig. 3b). When comparing the resulting pattern to the original one (Fig. 2c), we notice that the critical current is slightly suppressed and the side-lobes have somewhat random heights. We attribute these distortions and the shift from zero field to the trapping of magnetic flux in the superconducting film \cite{miller_85}. Indeed, the undistorted pattern seen in Figure 2c can be restored following the thermal cycling to $\approx 10$K, beyond the critical temperature of lead. 

\begin{figure}[ht]
\includegraphics[width=0.6 \columnwidth]{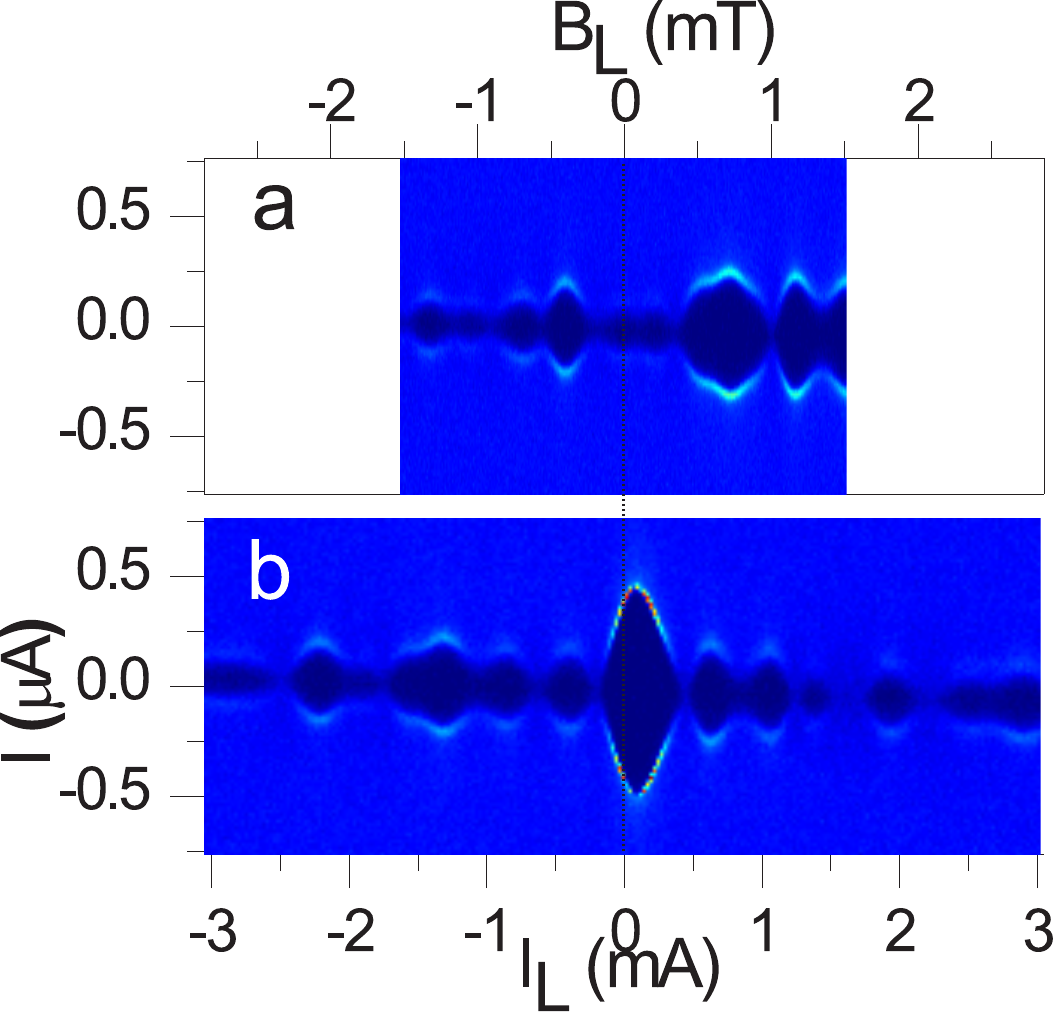}
\caption{\label{fig:overview}
 $R(I,I_L)$ measurements taken after the perpendicular magnetic field was ramped beyond several tens of mT. Panel (a) shows the measurement done when $B_{ext}$ was returned back to zero. Clearly, the pattern is now vastly distorted. Panel (b) is taken at $B_{ext}=3.4$mT. At this field, the original critical current modulation pattern (Fig. 2c) is partially restored. However, the central lobe shows a suppressed critical current, and the side-lobes form a distorted pattern. These permanent distortions are attributed to trapped flux in the Pb leads. Heating of the sample beyond the $T_C$ of Pb is required in order to restore the symmetric patterns seen in Figure 2c.
 }
\end{figure}

Interestingly, at fields less than those causing trapped flux of Figure 3, distortions of a different nature are introduced to the pattern (Fig. 2a). The first difference is that, returning to $B_{ext}=0$  restores the original pattern without any hysteresis. Second, the pattern demonstrates perfect symmetry under simultaneous reversal of both $B_{ext}$ and $I_{L}$, compare Figures 2 (a) and (e). We associate this behavior with the fact that the phase difference is not linear along the length of the leads. Indeed, the field $B_{L}$ may not be entirely uniform, so that it is not perfectly compensated by $B_{ext}$. Most likely, the deviations of $B_{L}$ from uniformity are caused by the bends in the leads (schematic in Fig. 1a), at which points the phase difference experiences discontinuous jumps proportional to $I_{L}$. The situation is very similar to the junctions with an artificial phase discontinuity controlled by an external current \cite{goldobin_sterck_2004,gaber_2005}. Indeed, some of the features we observe in Figures 2(a) and (e), \emph{e.g.} the strengthening of the side lobe at the expense of the central lobe, resemble those found in Refs. \cite{goldobin_sterck_2004,gaber_2005}. 

To describe the distortions found in Figure 2a,e, we consider a semi-realistic model of the sample. We assume that the leads extend from $x= -W/2$ to $+W/2$; the position-dependent phase difference $\Phi_{L}(x)$ (induced by $B_{L}(x)$) is taken to be piece-wise linear in $x$, with a slope proportional to $I_{L}$. Two identical discontinuous jumps of $\Phi_{L}(x)$ are placed at $-W/10$ and $+W/10$. The strength of the discontinuities is taken to be also proportional to $I_{L}$. These points are close to the actual locations of the bends in the leads, but we checked that the main features of the simulation do not crucially depend on the details (i.e. the position of discontinuities or symmetry of their placement). 

We also include the effect of the external field, presuming it induces uniform phase difference $\Phi_{ext}(x)$ along the length of the leads. In our simulations, the current-phase relation is assumed to be sinusoidal. Although deviations from a sinusoidal relation have been recently observed in SGS junctions \cite{Mason}, the approximation should be adequate in our case, due to the relatively large length between the leads ($L=400$ nm) and the relatively high temperature $\gtrsim 1$K. 

The simulated patterns of the critical current $I_C$ {\emph vs.} $I_{L}$ and $B_{ext}$ are shown in Figure 4. The main features observed in Figure 2 are qualitatively reproduced, such as: 1) the overall shift of the $I_C (I_{L})$ pattern in $B_{ext}$; 2) the growing distortion of the $I_C (I_{L})$ pattern in $B_{ext}$; 3) the growing strength of the side lobe on the high current side of the pattern at the expense of the central lobe; 4) the difference in width between that side lobe and the side lobes on the other side. Moreover, even the $B_{ext}=0$ curve, while similar to the perfect Fraunhofer pattern $I \propto \sin(\pi I_{L} / I_{L}^{(0)}) / I_{L}$, bears noticeable differences. Namely, some of the side lobes are suppressed almost to zero, while further lobes at higher $I_{L}$ regain strength. This type of behavior is indeed observed in experiment (Fig. 5). Note the region of suppressed critical current at $I_{L} \approx \pm2.3 mA$ in Figure 2c, and its reappearance at higher $I_{L}\approx \pm3.5 mA$. 

\begin{figure}[h]
\includegraphics[width=0.8 \columnwidth]{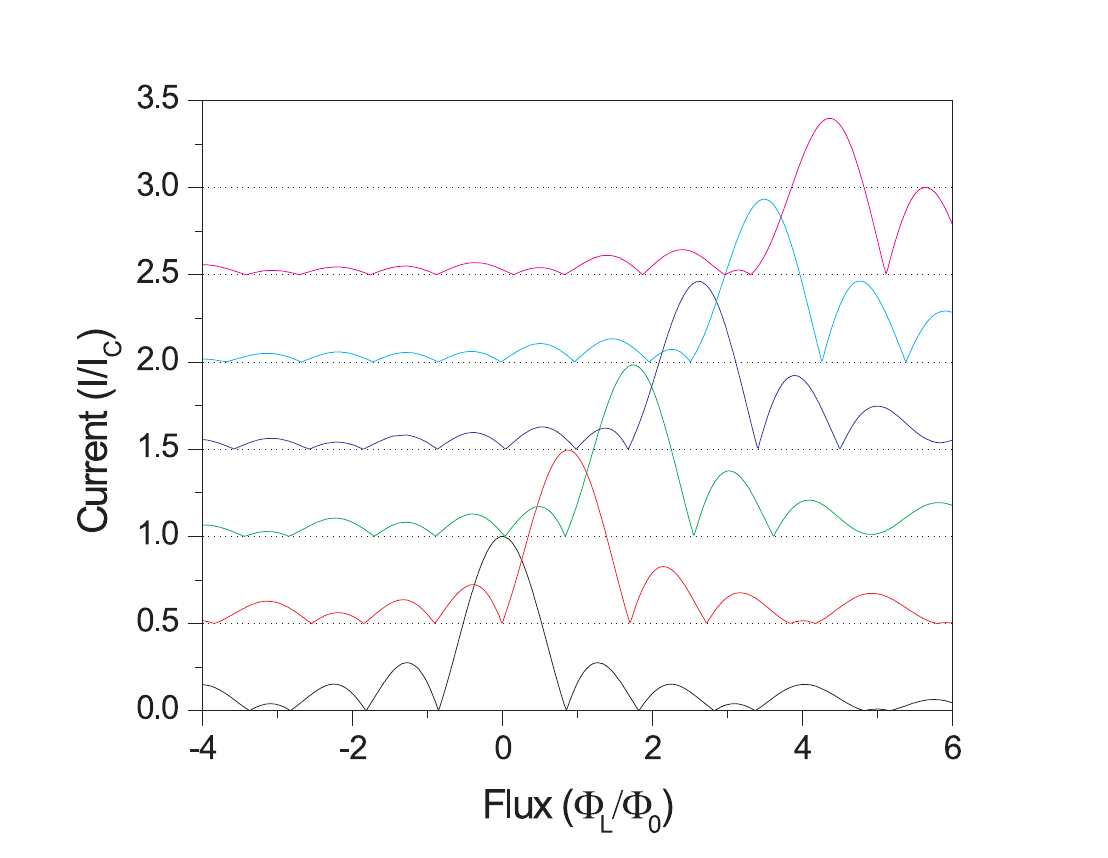}
\caption{\label{fig:overview}
Simulated critical current vs. $I_{L}$ at several values of $B_{ext}$. The bottom curve is taken at $B_{ext}=0$, and in each consecutive curvethe external flux grows in units of flux quantum $\Phi_0$. The phase difference between the two leads induced by $I_{L}$ is assumed to grow linearly along the length of the leads, proportionally to $I_{L}$, and to jump discontinuously at two locations by an amount also proportional to $I_{L}$.  This particular functional form it chosen to approximate the realistic shape of the sample, where the leads turn $90^{\circ}$ in two places; however the major features appear insensitive to the exact locations of the discontinuities. The horizontal axis is labeled in units of total flux $\Phi_L$ induced by $I_L$ excluding the discontinuities. The additional phase jump at each discontinuity is equal to $0.1 \pi \Phi_L / \Phi_0$.
}
\end{figure}

\begin{figure*}[ht]
\includegraphics[width=1.8 \columnwidth]{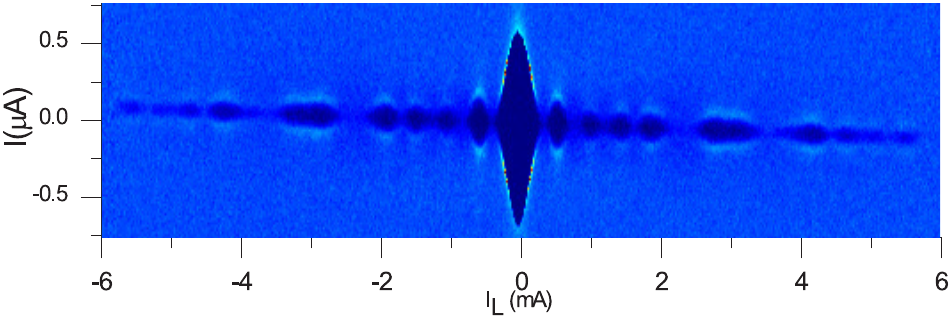}
\caption{\label{fig:overview}
Differential resistance $dV/d$I map measured {\emph vs.} bias current $I$ and current $I_{L}$ inducing magnetic field. This data are similar to Fig. 2c, but is extended up to $I_L=6mA$.  Note the  suppression of select side-lobes (see $I_L\approx \pm2.5mA$ and $\pm3.5mA$). This suppression of side-lobes qualitatively resembles the simulation results (Fig. 4, $B_{ext}=0$, $\Phi_L / \Phi_0\approx\pm3$ and $\pm 5$). The tilt of the pattern is an artifact of $I_L$ flowing through the normal part of the sample, thus creating a voltage drop that shifts the zero of $I$.
}
\end{figure*}

In conclusion, we demonstrate magnetic field-induced quasi-periodic modulation of critical current in Pb-graphene-Pb structures, which indicates their spatial uniformity. The magnetic field can be applied by running a current through one of the superconducting leads within the same structure, resulting in a simple, yet efficient method to {\it in situ} control the critical current. The dependence of the critical current on thus applied magnetic field deviates from the perfect Fraunhofer interference pattern. The difference is attributed to the presence of bends in the superconducting leads; a simple simulation supports this explanation.
  
The work was supported by the U.S. Department of Energy, Office of Basic Energy Sciences, Division of Materials Sciences and Engineering under Award DE-SC0002765.


\begin{thebibliography}{99}

\bibitem{heersche_2007} H. B. Heersche, P. Jarillo-Herrero, J. B. Oostinga, L. M. K. Vandersypen, and A. F.  Morpurgo, Nature {\bf 446}, 56 (2007).

\bibitem{miao_2007} F. Miao, S. Wijeratne, Y. Zhang,  U. C. Coskun, W. Bao, and C. N. Lau, Science {\bf 317}, 1530 (2007).

\bibitem{du_2008} X. Du, I. Skachko, and E. Y. Andrei, Phys. Rev. B {\bf 77}, 184507 (2008). 

\bibitem{gueron_2009} C. Ojeda-Aristizabal, M. Ferrier, S. Gueron, and H. Bouchiat, Phys. Rev. B {\bf 79}, 165436 (2009).

\bibitem{ivan}I. V. Borzenets, U. C. Coskun, S. J. Jones, and G. Finkelstein, Phys. Rev. Lett. {\bf107}, 137005 (2011)

\bibitem{huard_2008} B. Huard, N. Stander, J. A. Sulpizio, and D. Goldhaber-Gordon, Phys. Rev. B {\bf 78}, 121402 (2008).

\bibitem{avouris} F. Xia, V. Perebeinos, Y. Lin, Y. Wu, and P. Avouris, Nature Nano. {\bf 6}, 179 (2011).

\bibitem{novoselov_2005} K. S. Novoselov, D.  Jiang, F. Schedin, T.J. Booth, V. V. Khotkevich, S. V. Morozov, and A. K. Geim, PNAS {\bf 102}, 10451 (2005). 

\bibitem{Raman} A. C. Ferrari, J. C. Meyer, V. Scardaci, C. Casiraghi, M. Lazzeri, F. Mauri, S. Piscanec, D. Jiang, K. S. Novoselov, S. Roth, and A. K. Geim, Phys. Rev. Lett. {\bf 97} 187401, (2006).

\bibitem{tinkham_1996} M. Tinkham,  \emph{Introduction To Superconductivity} (McGraw-Hill, 1996).


\bibitem{miller_85} S. L. Miller, K. R. Biagi, J. R. Clem, and D. K.  Finnemore, Phys. Rev. B {\bf 31}, 268 (1985). 

\bibitem{gaber_2005} T. Gaber, E. Goldobin, A. Sterck, R. Kleiner, D. Koelle, M. Siegel, and M. Neuhaus, Phys. Rev. B {\bf 72}, 054522 (2005). 

\bibitem{goldobin_sterck_2004} E. Goldobin, A. Sterck, T. Gaber, D. Koelle, and R. Kleiner, Phys. Rev. Lett. {\bf 92}, 057005 (2004). 

\bibitem{Mason} C. Chialvo, I. C. Moraru, D. J. Van Harlingen, and N. Mason, arXiv:1005.2630v2 (2010)


\end{thebibliography}
\end{document}